\documentstyle[epsfig]{article}
\begin{document}

\def   \ni {\noindent}
\def   \cl {\centerline}
\def   \vs {\vskip}
\def   \hs {\hskip}
 
\def   \ssk {\vskip  5truept}
\def   \sk  {\vskip 10truept}
\def   \bsk {\vskip 15truept}
 
\def   \newpage {\vfill\eject}
\def   \newline {\hfil\break}
 
\def   \u {\underbar}

%

\hsize 5truein
\vsize 8truein
\font\abstract=cmr8
\font\keywords=cmr8
\font\caption=cmr8
\font\references=cmr8
\font\text=cmr10
\font\affiliation=cmssi10
\font\author=cmss10
\font\mc=cmss8
\font\title=cmssbx10 scaled\magstep2
\font\alcit=cmti7 scaled\magstephalf
\font\alcin=cmr6 
\font\ita=cmti8
\font\mma=cmr8
\def\ref{\par\noindent\hangindent 15pt}
\null


\title{\ni Kilohertz QPOs and strange stars
}                                               
\bsk \bsk
\author{\ni Tomasz Bulik, Dorota Gondek-Rosi\'nska, W{\l}odzimierz
Klu\'zniak$^1$ }                                                       
\bsk
\affiliation{Nicolaus Copernicus Astronomical Center, Bartycka 18,
00-716 Warszawa, Poland
\affiliation{~~~$^1$ and University of Wisconsin--Madison
}                                                
\bsk
\baselineskip = 12pt
\abstract{ABSTRACT \ni
The kilohertz quasi periodic oscillations (QPOs) discovered
in several low mass X-ray binaries (LMXBs)
by the Rossi X-ray Timing Explorer (XTE) are thought to occur at the
orbital frequency in accretion discs whose inner edge corresponds to
the innermost (marginally) stable orbit allowed by general
relativity. These ideas have been applied to constrain the equation of
state (e.o.s.) of the central neutron star. Here we discuss another
possibility, that the central object is a strange star,
and show how kHz QPOs constrain the e.o.s. of strange matter.
}
\bsk
\baselineskip = 12pt
\bsk
\baselineskip = 12pt
\text{\ni 1. INTRODUCTION}
\ssk
\ni
  
As first suggested by Bodmer (1971), bulk matter in its stable form
may be composed of deconfined up, down and strange quarks.  This
implies a possible existence of strange stars - compact objects
consisting of such quark matter (Witten 1984).
Glitching radio-pulsars are neutron stars and not strange stars
(Alpar 1987), and since their formation would be precluded in
a Galaxy contaminated by the disruption of a strange star in a binary merger
(Madsen 1988, Caldwell and Friedman 1991),
it has been suggested that strange stars cannot be formed directly in
supernovae but could exist as millisecond pulsars (Klu\'zniak 1994)
and be formed in LMXBs in an accretion-triggered phase transition
of neutron-star matter to strange matter (Cheng and Dai 1996). 
This phase transition could be accompanied by a gamma-ray burst
({\it ibid.}).

The launch of the XTE brought the discovery of high frequency QPOs in
the X-ray flux of bright Galactic sources (van der Klis {\it et~al.} 1996,
Strohmayer {\it et~al.} 1996).
Kilohertz QPOs have now been found in about a dozen LMXBs.  The QPOs
often come in frequency pairs, with the difference between the two
frequencies roughly constant for most objects. In some sources, for
example in 4U~1728-34 (Strohmayer {\it et al.} 1996), a third QPO
frequency equal to the difference between the two higher frequencies,
or its second harmonic, is observed during X-ray bursts. The usual
interpretation of these phenomena is that the third frequency is the
spin frequency of the star, whose beat with the highest frequency QPO
gives rise to the lower frequency of the ``kHz'' QPO pair. However, in
some sources the difference is not constant, e.g. in Sco~X-1
(van~der~Klis {\it et~al.} 1996) and in 4U1608-52
(M\'endez {\it et~al.} 1998), so the interpretation of the
lower frequency QPOs is uncertain.
\newpage

The X-ray flux of accreting degenerate stars may be modulated at the
orbital frequency (Bath 1973, Boyle, Fabian and Guilbert 1986) and it
has been expected that in the general-relativistic ``gap'' regime
(Klu\'zniak and Wagoner 1985)---in which the neutron star is inside
the marginally stable orbit, so the accretion disk cannot extend to
the stellar surface---clumps in the inner accretion disk may give rise
to modulations of flux in the kHz range with a characteristic maximum
frequency close to the orbital frequency in the marginaly stable orbit
(Klu\'zniak {\it et~al.} 1990).  There is no theory of QPOs, but in
keeping with this tradition we will identify the highest QPO frequency
with orbital frequency in the accretion disk.

In this work we assume that the compact objects in LMXBs are
strange stars and  discuss the implications of the
kilohertz QPOs on the properties of strange matter.
As we show below, the stringest lower limit on the density of bulk strange
matter corresponds to the highest orbital frequency.
The highest QPO frequency
so far has been observed in 4U~1636-53 at $1230\,$Hz 
(Zhang {\it et~al} 1997a).

\bsk
\ni 2. STRANGE MATTER EQUATION OF STATE
\ssk
\ni 

We describe strange matter by a stiff,
but still causal, equation of state:
\begin{equation}
P(\rho) = (\rho - \rho_0) \, c^2/3 \, ,
\end{equation}
where $\rho_0\equiv \rho_{14}\times10^{14}{\rm g/cm^3}$ 
is the density of bulk strange matter.
To determine the stellar mass-radius relation plotted in Fig. 1,
we solve the Oppenheimer-Volkoff equation (e.g. Shapiro and Teukolsky
1983). We note that the mass and radius of the star satisfy the scaling
relations (Witten 1984, Haensel {\it et al.} 1986):
\begin{equation}
{M(\rho_0')/M(\rho_0)} = \left({\rho_0/\rho_0'}\right)^{1/2},
~~~~~~~~
{R(\rho_0')/R(\rho_0)} = \left({\rho_0/\rho_0'}\right)^{1/2}
\label{scale}
\end{equation}
In the presence of a crust the relations~(2) are
approximate, however in the limit of large masses (close to the
maximal mass) they do hold.

\begin{figure}[t]
\centerline{\psfig{file=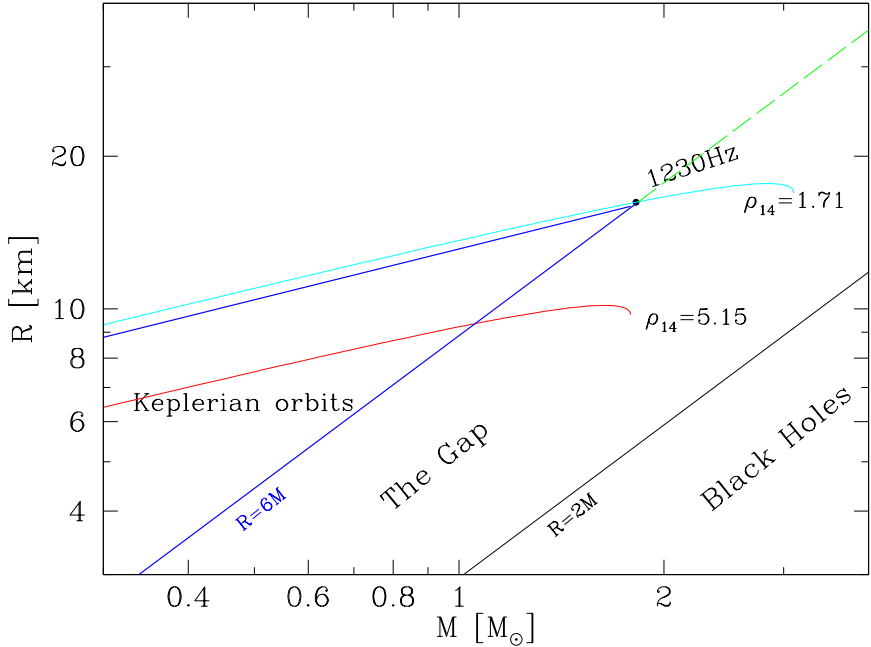, angle=-90, width=\textwidth}}
\caption{FIGURE 1. The mass radius diagram for non-rotating stars.
Keplerian orbits
are allowed in the region for which $R\ge 6M$. The upper thick line
corresponds to the keplerian frequency of $1230\,$Hz.
Mass radius relations for strange stars, corresponding to the limiting
cases described in the text, are also shown.
}
\end{figure}


\bsk
\ni 3. LIMITS ON $\rho_0$.
\ssk
\ni 

{\bf Lower limit}. Suppose that the highest QPO frequency observed,
1230~Hz, is a keplerian frequency.  The lines of constant keplerian
frequency, $f$, in a mass--radius diagram are described by
\begin{equation}
 R = \left( {2\pi f}\right)^{-2/3} (GM)^{1/3}\, .
 \end{equation}
They terminate at the radius of the marginally stable orbit,
$R=r_{ms} = 6 G M/c^2$. Since the star must fit inside the keplerian orbit,
the curve describing the stellar mass radius 
relation must intersect, or at least be tangent to, the wedge-shaped
region bounded from above by the line corresponding to the given 
keplerian frequency and 
by the marginally stable orbit line from below.
 This is shown in Figure 1, from which we see
that the observed frequency of
$1230\,$Hz implies a lower limit on the density of $\rho_0 >
1.71\times 10^{14}\,$g\,/cm$^3$.

{\bf Upper limit}. In order to find also an upper limit on the density
parameter, $\rho_0$, we make the additional assumption that the
maximum QPO frequency in a given source corresponds to orbital motion
in the marginally stable orbit (Kaaret {\it et al.} 1997, Klu\'zniak
1997, Zhang {\it et al.} 1997b).  This leads automatically to the
determination of the mass (for a non-rotating star):
\begin{equation}
M = 2.20 M_\odot  \times {1.00{\rm kHz}/ f} \, .
\end{equation}
Inserting $f=1230\,$Hz, we find that the mass
of the central object is $M = 1.78 M_\odot$.
This yields an upper limit on the strange matter density of
$\rho_0 <  5.15\times 10^{14}\,$g\,/cm$^3$. A~more stringent upper
bound is obtained by considering a lower frequency. There is strong evidence
that the QPO frequency saturates at the value $f=1060\,$Hz in the LMXB
4U 1820-30 (Zhang {\it et al.} 1998). Inserting this value of $f$ in eq. (4)
we obtain $M = 2.08M_\odot$. Eq. (2) then implies that the upper bound
on the density parameter decreases by the factor $(1060/1230)^2=0.743$,
yielding $\rho_0 <  3.82\times 10^{14}\,$g\,/cm$^3$. For
stars with moderate angular momentum $J$, the mass is increased by a factor
$1+0.75cJ/(GM^2)$ (Klu\'zniak {\it et al.} 1990), but deriving
a limit on $\rho_0$ would require construction of fully relativistic
models of rotating strange stars.
\newpage
%

\bsk
\ni 4. DISCUSSION
\ssk
\ni

We have shown that the existence of kilohertz QPOs leads to
constraints on the bulk strange matter density. For a slowly rotating
star (with angular momentum $J<<GM^2/c$) we find 
$ 1.7  \times 10^{14}\,$g\,/cm$^3\,<\rho_0 < 
3.8\times 10^{14}\,$g\,/cm$^3$. The only assumption used to
obtain the lower limit is that the maximum QPO frequency observed in LMXBs
corresponds to keplerian motion around a slowly rotating strange star.
To obtain the upper limit we assumed that the
highest observed QPO frequency in 4U 1820-30 corresponds to the orbital
frequency in the
marginally stable orbit. These limits can be improved by continued
observations of LMXBs, which may result in detection of higher QPO
frequencies in some sources and lower in others. The current upper limit
already rules out simplest models for the putative strange star in 4U 1820-30,
as the physical limits in a bag model with massless non-interacting
quarks are $4.2<\rho_{14}<6.5$
(Haensel, private communication).

\bsk
\baselineskip = 12pt
{\abstract \ni ACKNOWLEDGMENTS
This work has been supported in part by the KBN grants
2P03D00911,  2P03D01311, and made use of the NASA Astrophysics Data
System.
The authors thank Pawe{\l} Haensel for helpful comments.
}

\bsk
\baselineskip = 12pt


{\references \ni REFERENCES
\ssk

\def\apj{ApJ}
\def\apjl{ApJ. Lett. }
\def\iaucirc{IAUCirc}

\ref Alpar, A., 1987, Phys. Rev. Lett. 58, 2152

\ref Bodmer, A. R., 1971, Phys. Rev. 4, 1601

\ref Caldwell, R.R. and Friedman, J.~R., 1991, Physics. Lett. B 264, 143

\ref Cheng, K. S. and Dai, Z. G., 1996, Phys. Rev Lett. 77, 1210

\ref Haensel, P., Zdunik, J. L., Schaefer, R., 1986 A\&A, 160, 121

\ref Kaaret, Ph., Ford, E.C. and Chen, K. 1997, \apj, 480, L27

\ref {Klu\'zniak}, W., 1994, A\&A 286, L17

\ref {Klu\'zniak}, W., 1997, astro-ph/9712243

\ref {Klu\'zniak}, W. and {Wagoner}, R.~V., 1985, \apj, 297, 548

\ref Klu\'zniak, W., Michelson, P. and Wagoner, R.~V. 1990, \apj, 358, 538

\ref Madsen, J., 1988, Phys. Rev. Lett. 61, 2909

\ref M\'endez, M. et~al. 1998, ApJ 505, L23

\ref Shapiro, S, and Teukolsky, S. A., 1983, Black Holes, White
Dwarfs, and Neutron Stars (New York: Wiley)

\ref{Strohmayer}, T.~E., {Zhang}, W., and {Swank}, J.~H., {Smale}, A., 
  {Titarchuk}, L., {Day}, C. and {Lee}, U. 1996, \apj, 469, L9

\ref {Van Der Klis}, M., {Swank}, J., {Zhang}, W., {Jahoda}, K.,
 {Morgan}, E.~H., {Lewin}, W.~H.~G., {Vaughan}, B., and {Van Paradijs}, J.,
 1996, \apj, 469, L1

\ref Witten E. 1984, Phys. Rev. D., 30, 272

\ref {Zhang}, W., {Lapidus}, I., {Swank}, J.~H., {White}, N.~E.,
 and {Titarchuk},
  L., 1997{{a}}, \iaucirc, 6541

\ref
{Zhang}, W., {Strohmayer}, T.~E., and {Swank}, J.~H., 1997{{b}},
  \apj, 482, L167

\ref
{Zhang}, W., Smale, A.~P., {Strohmayer}, T.~E., and {Swank}, J.~H., 1998,
 \apj, 500, L171
}                      

\end{document}